\documentclass[%
 reprint,
superscriptaddress,
 amsmath,amssymb,
 aps,
floatfix,
twocolumn,
]{revtex4-2}
\usepackage{graphicx}
\usepackage{dcolumn}
\usepackage{bm}
\usepackage{adjustbox}
\usepackage{xcolor}
\usepackage{amsmath}
\usepackage{nccmath}
\usepackage{ulem}

\newcommand{\+}{\dagger}

\newcommand{\f}{\frac}
\newcommand{\8}{\infty}
\newcommand{\ket}{\rangle}
\newcommand{\bra}{\langle}

\newcommand{\romann}[1]{\uppercase\expandafter{\romannumeral#1}}

\begin{document}


\title{Strong Coupling Quantum Thermodynamics far away from Equilibrium: \\
Non-Markovian Transient Quantum Heat and Work}

\author{Wei-Ming Huang}
\affiliation{Department of Physics and Center for Quantum Information Science, National Cheng Kung University, Tainan 70101, Taiwan }

\author{Wei-Min Zhang}%
\email{ Correspondence author: wzhang@mail.ncku.edu.tw}
\affiliation{%
Department of Physics and Center for Quantum Information Science, National Cheng Kung University, Tainan 70101, Taiwan
}
\affiliation{Physics Division, National Center for Theoretical Sciences, Taipei 10617, Taiwan}

\date{\today}

\begin{abstract}
In this paper, we investigate the strong coupling quantum thermodynamics of the hybrid quantum system far away from equilibrium. The strong coupling hybrid system consists of a cavity and a spin ensemble of the NV centers in diamond under external driving that has been realized experimentally. We apply the renormalization theory of quantum thermodynamics we developed recently to study the transient quantum heat and work in this hybrid system. We find that the dissipation and fluctuation dynamics of the system induce the transient quantum heat current which involve the significant non-Markovian effects. On the other hand, the energy renormalization and the external driving induce the quantum work power. The driving-induced work power also manifests non-Markovian effects due to the feedback of non-Markovian dynamics of the cavity due to its strong coupling with the spin ensemble.
\end{abstract}

\pacs{Valid PACS appear here}
\maketitle

\section{introduction}
The investigation of quantum thermodynamics far away from the equilibrium has attracted a great attention in the last decade \cite{Vinjanampathy2016,Millen2016,Landi2021,Huang2022,Jarzynski2011,Kosloff2013,Xiong2015,Nandkishore2015,Binder2018,Deffner2019,Talkner2020}. Thermodynamics of nanoscale systems can exhibit exotic properties. For example, quantum coherence and entanglement could enhance the efficiency of heat engines in comparison with classical counterparts \cite{Haack2019,Scully2011,Uzdin2015,Rosnagel2014,Bergenfeldt2014,Zhang2014,Rosnagel2016}. It was also argued that quantum interference boosts the conversion efficiency in photosynthesis as a quantum engine \cite{Dorfmana2013,Engel2007,Scholes2011,Creatore2013,Duan2017,Harush2021}. However, these exotic quantum thermodynamics properties are usually extracted from the systems weakly coupled with their reservoirs. For the strong coupling between system and reservoir, the energy conversion of heat and work is not clearly understood \cite{Talkner2020,Hsiang2018}. One of main motivations in the study of quantum thermodynamics is to understand and manipulate energy conversion in nano and atomic scale quantum systems when they strongly couple to their environment.

In fact, the definitions of thermodynamics quantities, such as heat and work, are quite ambiguous at quantum level. A contradiction  has been pointed out in evaluating specific heat in strong coupling system due to the different definition of internal energy \cite{Hanggi2008,Ingold2009}. The definition of heat and work depend on how to take into account properly the coupling energy between the system and its reservoir \cite{Seifert2016,Esposito2015,Perarnau-Llobet2018,Esposito2010}. Even for a system operating at steady-state limit, the definition of heat and work are still debated within the framework of quantum mechanics \cite{Nicolin2011,Gaspard2015,Topp2015,Dhar2012}. For the system far away from equilibrium, the transient processes of energy exchange becomes much more complicated in the strong coupling regime. In our previous work, we have developed a renormalization theory of nonequilibrium quantum thermodynamics for both the weak to strong couplings \cite{Huang2022}. In this theory, we provides a rather unambiguous definitions of quantum thermodynamics quantities, including quantum heat and work. In this paper, we shall study the transient energy exchange in nonequilibrium quantum thermodynamics.

We shall focus on a realistic system, a microwave superconducting cavity strongly interacted with a spin ensemble made of NV centers in diamond. This system has been experimentally realized on its non-Markovian decoherence and quantum memory \cite{Putz2014,Putz2017}. Here, we shall study its nonequilibrium quantum thermodynamics based on the exact master equation. The dynamics of heat and work in this system is explored. We find that the non-Markovian dynamics induce the energy exchanges through quantum heat and work under the external driving. We also find that thermal fluctuations produce quantum heat current. Finally, a physical picture of transient energy exchange in quantum thermodynamics is obtained.

The rest of this paper is organized as follows.
In Sec.~\ref{NEQ}, we present a generalized Tavis-Cummings model for the hybrid system of the microwave cavity coupling strongly to the spin ensemble and the nonequilibrium theory for such a strongly coupling system. In Sec.~\ref{QT}, we apply the quantum thermodynamics based on the nonequilibrium theory to study the transient quantum heat and work in the strong coupling regime. We find that the dissipation and fluctuation dynamics of the system induce transient quantum heat current which involves strong non-Markovian effects. On the other hand, the quantum work power arises from external drivings and the energy renormalization. Through the renormalization of the cavity filed, the quantum work manifests non-Markovian dynamics. By tuning the driving and cavity frequency as well as the coupling strength between the cavity and spin ensemble, we show how the system non-Markovian dynamics affects the energy exchange in a strong coupling system.
A conclusion is drawn in Sec.~\ref{conclusion}.

\section{nonequilibrium theory of a strong coupling hybrid system}\label{NEQ}
The strong coupling cavity system concerned in this paper is investigated experimentally by Putz, {\it et. al}, \cite{Putz2014,Putz2017}. It is a superconducting microwave cavity strongly coupled to a spin ensemble of the NV centers in diamond. Also, the cavity is driven by an external pulse so that one can manipulate and measure the nonequilibrium photon dynamics. Theoretically, the cavity photon dynamics can be well described by the generalized Tavis-Cummings model with the following Hamiltonian \cite{Tavis1968},
\begin{align} \label{csH}
  H(t)\!&=\!\hbar \omega_ca^\+a\!+\![f(t)a^\+\!+\!\text{h.c.}]\!+\!\sum_i\hbar\Delta_i\sigma^z_i\!+\!\sum_k\hbar \omega_kb^\+_kb_k\notag\\
  &\!+\!\sum_i(V_ia^\+\sigma^-_i\!+\!V^*_i\sigma^+_ia)
  \!+\!\sum_k(V_ka^\+b_k\!+\!V^*_kb_k^\+a).
\end{align}
Here $a^\+(a)$ is the creation (annihilation) operator of cavity photon mode $\omega_c$, and $f(t)$ is the external driving field applied to the cavity. The operator 
$\sigma^z_i$,$\sigma^\pm_i$ represent the three Pauli matrices of the $i$-th spin in the spin ensemble with spin energy level splitting $\hbar\Delta_i$. The parameter $V_i$ is the coupling strength between the cavity mode and the $i$-th spin of the spin ensemble. We also include the cavity leakage effects in Eq.~(\ref{csH}) due to the weak coupling between the cavity and the free space electro-magnetic (EM) modes $\omega_k$, where $b^\+_l(b_l)$ is the corresponding creation (annihilation) operator and $V_k$ is the coupling strength between the cavity and EM modes.
\begin{figure}[h]
  \includegraphics[width=8cm, height=3cm]{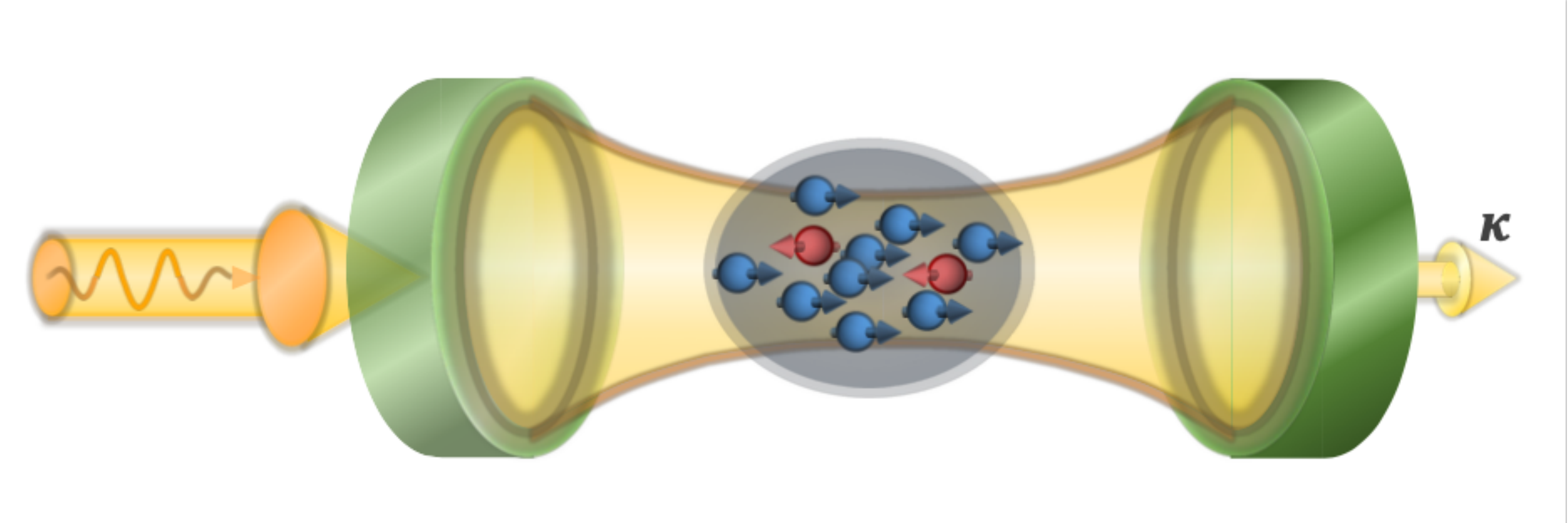}
  \caption{A schematic diagram of the strong-coupling hybrid system consisting of a superconducting microwave cavity coupled with a spin ensemble of NV centers in diamond.}
\end{figure}

In the practical experimental setup, spins of the NV centers in diamond are surrounded by Helmholtz coil which supplies a strong magnetic field to modify all the spins into the polarized ground state. The total spin number is the order of $10^{12}$. The external driving field applying to the cavity can excite about a small number of spins $(\approx 10^6)$ \cite{Putz2014,Putz2017}. This implies that the spin ensemble is a highly polarized spin ensemble. Thus, the Holstein-Primakoff approximation \cite{Primakoff1939}, $\sigma^z_i \equiv c^\+_ic_i -1/2$ and 
$\sigma^\pm_i \equiv c^\+_i (1-c^\+_ic_i)^{-1/2} \simeq c^\+_i$, can be applied to bosonlize the spin ensemble, where $c_i^\+(c_i)$ represents the bosonic creation (annihilation) operator of the corresponding $i$-th spin. As a result, the Hamiltonian~(\ref{csH}) can be reduced to
\begin{align} \label{H}
  H(t)\!&=\!\hbar \omega_ca^\+a\!+\![f(t)a^\+\!+\!\text{h.c.}]\!+\!\sum_i\hbar\Delta_ic^\+_ic_i\!+\!\sum_k\hbar \omega_kb^\+_kb_k\notag\\
  &\!+\!\sum_i(V_ia^\+c_i+V^*_ic^\+_ia)
  +\sum_k(V_ka^\+b_k+V^*_kb_k^\+a).
\end{align}

The nonequilibrium theory of this system has been formulated \cite{Lei2012,Chiang2021}. The cavity photon dynamics can be described by the cavity density matrix $\rho_c(t)={\rm Tr_E}[\rho_{tot}(t)]$. The total density matrix $\rho_{tot}(t)$ is determined from the Liouville-von Neumann equation of the total system (i.e., the cavity plus the spin ensemble and the free space EM modes), 
\begin{align}
  i\hbar\f{d}{dt}\rho_{tot}(t)=[H(t),\rho_{tot}(t)].
\end{align}
After tracing over all the environment degrees of freedom (including the spin ensemble and the free space EM modes), we obtain the exact master equation of the cavity photonic state \cite{Lei2012,Chiang2021},
\begin{align}
  \label{master}
  \f{d}{dt}&\rho_c(t)=
  \f{1}{i\hbar}[H^r_c(t,t_0),\rho_c(t)]\notag\\
  &+\gamma(t,t_0)[2a\rho_c(t)a^\+\!-\!a^\+a\rho_c(t)\!-\!\rho_c(t)a^\+a]\\
  &+\tilde{\gamma}(t,t_0)[a\rho_c(t)a^\+\!+\!a^\+\rho_c(t)a\!-\!a^\+a\rho_c(t)\!-\!\rho_c(t)aa^\+].\notag
\end{align}
Here, the first term describes the unitary evolution of the cavity density matrix with the renormalized Hamiltonian 
\begin{align}
  H^r_c(t,t_0)=\hbar \omega^r_c(t,t_0)a^\+a+f_r^*(t,t_0)a+f_r(t,t_0)a^\+.
\end{align} 
The renormalization is given by the renormalized frequency $w^r_c(t,t_0)$ and the renormalized driving field $f_r(t,t_0)$ arisen from the cavity coupling to the spin ensemble. The second and third terms in Eq.~(\ref{master}) give the non-unitary evolution of the cavity induced by the spin ensemble and also the leakage effect, the coefficients $\gamma(t,t_0)$ and $\tilde{\gamma}(t,t_0)$ describe dissipation and fluctuations, respectively. They characterize the cavity spontaneous emission into the spin ensemble (including the free space leakage) and the induced cavity emission and absorption from the thermal fluctuations of the spin ensemble and the free space.

All the time-dependent parameters in the exact master equation (\ref{master}), i.e., the renormalized frequency and the renormalized driving field, as well as the dissipation and fluctuation coefficients are nonperturbatively and exactly determined by the nonequilibrium Green functions \cite{Lei2012},
\begin{subequations}\label{coeff_MEQ}
\begin{align}
  &i\omega^r_c(t,t_0)+\gamma(t,t_0)=-\f{\dot{u}(t,t_0)}{u(t,t_0)},\\
  &f_r(t,t_0)=i\hbar\dot{y}(t,t_0)-i\hbar[\f{\dot{u}(t,t_0)}{u(t,t_0)}y(t,t_0)], \label{f}\\
  &\tilde{\gamma}(t,t_0)=\dot{v}(t,t)-[\f{\dot{u}(t,t_0)}{u(t,t_0)}v(t,t)+c.c].
\end{align}
\end{subequations}
The Green functions $u(t,t_0)$, $v(\tau,t)$ and the driving-induced cavity field $y(t,t_0)$ are determined non-perturbatively by the following equations,
\begin{subequations}\label{equyv}
\begin{align}
  &\f{d}{dt}u(t,t_0)+i\omega_cu(t,t_0)+\int^t_{t_0}\!d\tau g(t,\tau)u(\tau,t_0)=0,\\
  &y(t,t_0)=\f{1}{i\hbar}\int^t_{t_0}\!d\tau u(t,\tau)f(\tau),\label{y}\\
  &v(\tau,t)=\int^\tau_{t_0}\!dt_1\int^t_{t_0}\!dt_2 u(\tau,t_1)\tilde{g}(t_1,t_2)u^*(t,t_2).\label{v}
\end{align}
\end{subequations}
Here, the integral kernels,
\begin{subequations}
\begin{align}
  g(t,\tau)=&\int_0^\8\f{d\omega}{2\pi}J(\omega)e^{-i\omega(t-\tau)},\\ 
  \tilde{g}(t,\tau)=&\int_0^\8\f{d\omega}{2\pi}J(\omega)\bar{n}(\omega,T)e^{-i\omega(t-\tau)},
\end{align}
\end{subequations} 
are the time correlation functions between the cavity and the spin ensemble (also include the free space for the leakage). The non-Markovian memory due to the back-reaction between the cavity and environment is described by the time-convolution equation (\ref{equyv}). The spectrum density of the environment (spin ensemble plus the free space) 
\begin{align}
J(\omega)=&\f{2\pi}{\hbar^2}\bigg[\sum_i |V_i|^2\delta(\omega-\Delta_i/\hbar)\!+\!\sum_k |V_k|^2\delta(\omega-\omega_k)\bigg]\notag\\
=&J_s(\omega)+J_e(\omega)
\end{align}
represents the spectrum densities of the spin ensemble and the free space EM field coupled with the cavity. The particle distribution $\bar{n}(\omega,T_0)=1/(e^{\hbar \omega/k_BT_0}-1)$ is the distribution of boson mode $\omega$ with temperature $T_0$ at the initial time $t_0$.

The spectrum density of the spin ensemble characterizes the inhomogeneous spectrum broadening due to the local magnetic dipole-dipole couplings between NV centers and the residual nitrogen paramagnetic impurities. It can be measured and manipulated in experiments \cite{Putz2017}. It was found in the experiment \cite{Putz2014} that the spectrum density of the spin ensemble $J_s(\omega)$ is a q-Gaussian spectrum, which is an intermediate form between a Gaussian spectral density (q = 1) and a Lorentzian spectral density (q = 2). 
\begin{align}\label{spectrum}
  J_s(\omega)&=2\pi\Omega^2C[1-(1-q)\f{(\omega-\omega_s)^2}{\Delta^2}]^\f{1}{1-q}
\end{align}
where $q=1.39$ is experimentally fitted, $C$ is a normalization constant of the density of state, $\omega_s$ is the main frequency of the spin ensemble,  $\Delta$ is determined by the full-width at the half maximum of $J_s(\omega)$ which is given by $d=2\Delta\sqrt{\f{2^q-2}{2q-2}}$, $\Omega$ is the coupling strength and $\Omega\sim d$ represents a strong coupling.
The free EM spectrum $J_e(\omega)$ is taken by the decay constant $\kappa$, describing the cavity leakage. Thus, $J(\omega)=J_s(\omega)+2\kappa$. 
  
In the experimental setup \cite{Putz2014,Putz2017}, the main frequency of spin ensemble is resonant with the cavity frequency, $\omega_s=\omega_c=2\pi\times2.69$ GHz, the half-width of the spin spectrum is $d=18.8\pi$ MHz. The coupling strength $\Omega$ is experimentally variable. Also notice that although the spin ensemble is surrounded by a Helmholtz coil so that a strong magnetic field can make all the spins into the polarized ground state, the system cannot be in the absolute zero temperature. To reduce the thermal fluctuation effect, the original experimental setup is cooled and the whole system down to $T=25$ mK, which is only about one-fifth of the excitation energy of spins. Here, to study the thermodynamics of this hybrid system, we make the temperature and cavity frequency variable, i.e., experimentally adjustable. With the above nonequilibrium theory based on the exact master equation, we are now able to explore the nonequilibrium quantum thermodynamics of this hybrid system under the external deriving field.

\section{The transient quantum work and heat}\label{QT}
\subsection{The general definition of quantum heat and quantum work}
In the convention thermodynamics, the internal energy of a system is solely determined by the system Hamiltonian where it is assumed that the system interacts only very weakly with the reservoir. Beyond the weak coupling regime, however, the internal energy must take system-environment coupling energy into account. In the literature, this is a difficult problem that has not been solved because it is not clean how much coupling energy should be included in the system's Hamiltonian \cite{Hanggi2008,Ingold2009,Seifert2016,Esposito2015,Perarnau-Llobet2018}. Based on our exact master equation, this difficult problem has been unambiguously overcome in our recent theory of renormalized quantum thermodynamics from weak to strong couplings \cite{Huang2022}.
In this renormalized quantum thermodynamics theory, the modification of the system Hamiltonian by the system-reservoir coupling is given by the renormalization of the Hamiltonian in the exact master equation Eq.~(\ref{master}). We thus define the internal energy as the average of the renormalized Hamiltonian in Eq.~(\ref{master}),
\begin{subequations}
\begin{align}
  E^r(t)\equiv&\bra H^r_c(t)\ket=Tr_{_S}[H^r_c(t)\rho_c(t)].\label{E}
\end{align}
\end{subequations}
Thermodynamically, energy can enter into or leave from a system through heat and work. Heat is arisen from the change of the entropy in state populations. Quantum mechanically, work corresponds to the changes in energy levels of the system and also the driving field. Accordingly, we have introduced the quantum work power and quantum heat current as follows  \cite{Huang2022}.
\begin{subequations}\label{defWQ}
  \begin{align}
      \mathcal{P}_w(t) &\equiv Tr_{_S}\bigg[\f{dH^r_c(t)}{dt}\rho_c(t)\bigg],\\
      \mathcal{I}_h(t) &\equiv Tr_{_S}\bigg[H^r_c(t)\f{d\rho_c(t)}{dt}\bigg].
  \end{align}
\end{subequations}

Explicitly, based on the exact master equation (\ref{master}), the above transient work power and the heat current can be expressed explicitly as

\begin{subequations}\label{WQ_MEQ}
  \begin{align}
    &\mathcal{P}_w(t)\!=\!
    \bar{n}(t)\f{d}{dt}\bigg[\hbar \omega^r_c(t)\bigg]+2\text{Re}\bigg[\bra a(t)\ket\f{df^*_r(t)}{dt}\bigg],\\
    &\mathcal{I}_h(t)\!=\!
        \hbar \omega^r_c(t)\tilde{\gamma}(t)\!-\!2\gamma(t)\bigg[\!E^r(t)\!-\!\text{Re}[\bra f^*_ra\ket(t)]\!\bigg].\label{Q_MEQ}
  \end{align}
\end{subequations}
For simplicity, here we set $t_0=0$. In Eq.~(\ref{WQ_MEQ}a), the quantum work power consists of two parts, an intrinsic part and an extrinsic part. The intrinsic part is $\mathcal{P}_w^e=\bar{n}(t)d[\hbar \omega^r_c(t)]/dt$, which corresponds to the energy renormalization due to the coupling to the environment. As the system interact with its environment, the system and the environment do the work on each other through the shift of their energy levels. In addition to the intrinsic work done on the system, the extrinsic work is given by the driving-induced system work power $\mathcal{P}_w^d(t)=2\text{Re}[\bra a(t)\ket df^*_r(t)/dt]$ where the renormalization of the driving field is induced by the backaction of the non-Markovian cavity dynamics, as shown in Eqs.~(\ref{coeff_MEQ}b) and (\ref{equyv}b).

On the other hand, the quantum heat current in Eq.~(\ref{Q_MEQ}) consists of two terms: $\hbar \omega^r_c(t)\tilde{\gamma}(t)$ and $-2\gamma(t) \big[E^r(t)\!-\!\text{Re}[\bra f^*_ra\ket(t)]\big].$ They are proportional to the fluctuation coefficient $\tilde{\gamma}(t)$ and the dissipation coefficient $\gamma(t)$, respectively, see Eqs.~(\ref{master}) and (\ref{coeff_MEQ}). It shows that quantum heat can be induced not only from fluctuation but also from dissipation dynamics. The internal energy consists of the initially related energy and the energy arisen from thermal fluctuations, i.e., $E^r(t)=E^r(t)|_{_{T=0}}+\hbar \omega_c^rv(t,t)$, where $v(t,t)$ is the thermal fluctuation correlation. Thus, we can define respectively the fluctuation heat current $\mathcal{I}_{h}^{\mathcal{F}}(t)$ and the dissipation heat current $\mathcal{I}_{h}^{\mathcal{D}}(t)$ as follows: 
\begin{subequations}
  \label{heat_currents}
\begin{align}
  \mathcal{I}_{h}^{\mathcal{F}}(t)&\!=\!\hbar \omega^r_c(t)\!\big[\tilde{\gamma}(t)\!-\!2\gamma(t)v(t,t)\!\big]\!=\!\hbar w^r_c(t) \dot{v}(t,t),\\
  \mathcal{I}_{h}^{\mathcal{D}}(t)&\!=\!-2\gamma(t) \!\big[E^r(t)|_{_{T=0}}\!-\!\text{Re}[\bra f^*_ra\ket(t)]\big].
\end{align}
\end{subequations}
It shows that the fluctuation heat current $\mathcal{I}_{h}^{\mathcal{F}}(t)$ is fully determined by the thermal fluctuations correlation $v(t,t)$. The dissipation heat current $\mathcal{I}_{h}^{\mathcal{D}}(t)$ is purely governed by the dissipation dynamics. Moreover, Eq.~(\ref{heat_currents}) are valid for both the weak and strong couplings. With these results, we can investigate the transient energy exchange in strong coupling systems.  
  
\subsection{The transient quantum work and heat at strong coupling}
We consider first a simple situation of the cavity being initially in a coherent state $|z_0\ket$ and decoupled with the reservoir $(J(\omega)=0)$ to understand the pure driving dynamics. The cavity is driven by an external time-dependent field $f(t)=Ae^{-i\omega_d t}$ with the constant amplitude $A$ and the frequency $\omega_d$. In this trivial case, we obtain the free Green function $u(t,\tau)=e^{-i\omega_c(t-\tau)}$ from Eq.~(\ref{equyv}) for the cavity. There is no energy renormalization to the system because the cavity is decoupled from the reservoir. Also, the driving field $f_r(t)=f(t)$ remains unchanged. The cavity photon state evolve into a coherent sate $|\psi(t)\ket=\exp\{z(t)a^\+ +z^*(t)a\}|0\ket$ where $z(t)=[z_0+\f{A}{\hbar \delta}(1-e^{i\delta t})]e^{-i\omega_c t}$ is the field amplitude and $\delta=\omega_c-\omega_d$ is the detuning. From these results, we obtain the average energy of cavity and the cavity work power driving by the external field,
\begin{subequations}\label{isolated}
  \begin{align}
    E(t)=&~\bra H_c\ket(t)\notag\\
    =&~\hbar \omega_c |z_0|^2+2\text{Re}[\f{A^*z_0}{\delta}(w_c-w_de^{-i\delta t})]\notag\\
    &~+\f{2|A|^2}{\hbar\delta^2}\omega_d\big[1-\cos(\delta t)\big],\\
    \mathcal{P}_w(t)
    =&~2\text{Re}[z(t)\f{df^*(t)}{dt}]\notag\\ 
    =&~\f{2|A|^2}{\hbar \delta}\omega_d \sin(\delta t)-2\text{Im}[A^*z_0e^{-i\delta t}]w_d,\\
    \mathcal{I}_h(t)=&~0.
  \end{align}
\end{subequations} 
Of course, there is no heat production in an isolated system as one expected. By tuning the driving field frequency, we can control the power dynamics and drive the cavity system cyclically as an ideal optical engine. Furthermore, Eqs.~(\ref{isolated}) is reduced to $E(t)=\hbar \omega_c|z_0|^2+\f{|A|^2}{\hbar}\omega_ct^2$, $\mathcal{P}_w(t)=\f{2|A|^2}{\hbar}\omega_c t$ if we tune the driving field to the resonance and in phase with the system, i.e. $\delta=0$ and $\text{Im}[A^*z_0]=0$. Thus, the cavity behaves as an ideal optical resonator.

\subsubsection{The quantum work power and the quantum heat currents}
We now return to the hybrid cavity system. We first consider that the spin ensemble is at a finite temperature $T_0$, and meantime we also turn off the driving. Because the  spectrum density of the spin ensemble is symmetric with respect to the cavity frequency $\omega_c$, i.e., $\omega_s=\omega_c$ in Eq.~(\ref{spectrum}). We find that there is no cavity frequency shift by the cavity-spin coupling due to the symmetric properties of spectral density with respect to the cavity frequency. In this case, there is no energy renormalization so that the quantum work is zero. As a result, the changes of internal energy are caused only by the dissipation heat and the fluctuation heat, i.e., $dE^r(t)/dt=\mathcal{I}_h^{\mathcal{D}}(t)+\mathcal{I}_h^{\mathcal{F}}(t)$ in the nonequilibrium evolution. 

\begin{figure}[h]
  \includegraphics[width=8.4cm, height=6cm]{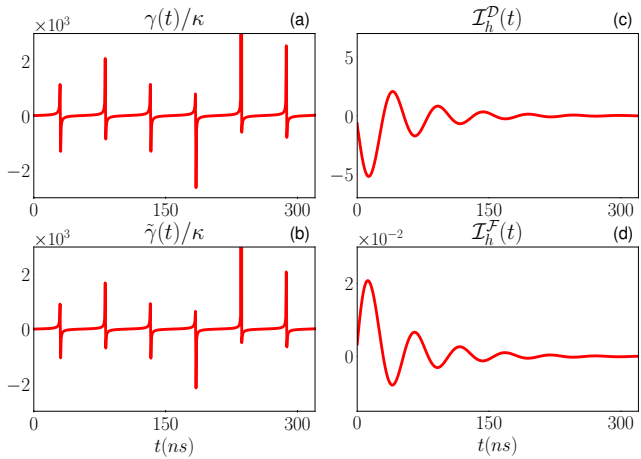}
  \caption{The cavity is strongly coupled to the spin ensemble with main frequency $\omega_s=\omega_c$. (a) The dissipation coefficient $\gamma(t)$, (b) the fluctuation coefficient $\tilde{\gamma}(t)$, (c) the dissipation heat current $\mathcal{I}_{h}^{\mathcal{D}}(t)$, and (d) the fluctuation heat current $\mathcal{I}_{h}^{\mathcal{F}}(t)$ are plotted as a function of time. In (c,d), the units are $\hbar \omega_c/ns$. The system is initially prepared in a coherent state $|z_0\ket$ with $z_0=10$. Other parameters are the coupling strength $\Omega=17.2\pi$ MHz,  the decay constant $\kappa=0.8\pi$ MHz, and the temperature $T_0=0.1$~K. No driving is applied in this case.} \label{resonance}
\end{figure}

In Fig.~\ref{resonance}, we show the dissipation coefficient $\gamma(t)$ and the fluctuation coefficient $\tilde{\gamma}(t)$ in the left panel, and in the right panel, the plots are the dissipation heat current $\mathcal{I}_{h}^{\mathcal{D}}(t)$ and the fluctuation heat current $\mathcal{I}_{h}^{\mathcal{F}}(t)$ as a function of time. The strong oscillations of dissipation and fluctuation coefficients between positive and negative values shown in Fig.~\ref{resonance}(a,b) manifest the strong non-Markovian dynamics in the strong coupling. Due to the strong coupling, the system and the environment rapidly exchange information each other, as the memory effect. Such rapid oscillations accompany system information fast flowing into and out of the spin ensemble, describing entropy production, the source of heat. Thermodynamically, the non-Markovian entropy production generates flowing-back heat currents, as shown in Fig.~\ref{resonance}(c,d). Figure~\ref{resonance}(c) and (d) also show that the dissipation heat current is always out of phase with the fluctuation heat current. This out of phase phenomenon indicates that the dissipation and fluctuation dynamics, both bring the energy and fluctuation flowing back and forward between the system and environment, will eventually make the system and environment approach thermal equilibrium.

\begin{figure}[h]
  \includegraphics[width=8.4cm, height=6cm]{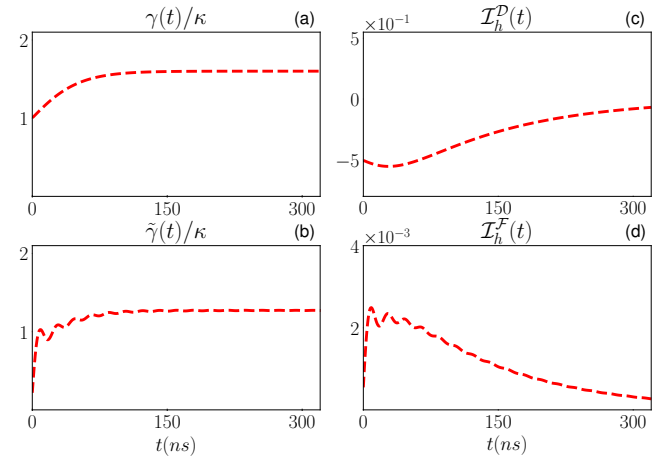}
  \caption{The cavity is weakly coupled to the spin ensemble with main frequency $\omega_s=\omega_c$. (a) the dissipation coefficient $\gamma(t)$, (b) the fluctuation coefficient $\tilde{\gamma}(t)$, (c) the dissipation heat current $\mathcal{I}_{h}^{\mathcal{D}}(t)$, and (d) the fluctuation heat current $\mathcal{I}_{h}^{\mathcal{F}}(t)$ are plotted as a function of time. The coupling strength $\Omega$ is $1.72\pi$ MHz. Other parameters and units are the same in Fig.~\ref{resonance}}  
  \label{resonance_weak}
\end{figure}

As a comparison, we also show the results in the weak coupling regime \cite{Putz2014}, see Fig.~\ref{resonance_weak}. In contrast to the strong coupling regime, the dissipation and fluctuation coefficients in Fig.~\ref{resonance_weak}(a,b) monotonically approach to a steady value, i.e., there are no oscillations.  As a result, the dissipation heat current is always negative, carrying energy and information left away from the system, as shown by Fig.~\ref{resonance_weak}(c,d). However, the fluctuation heat current is always positive, namely it brings energy and information from the environment back into the system. These single-direction energy and information transfers are typical Markovian processes in the weak coupling regime.

To show a more complete picture of energy and information exchange between the system and environment, we next discuss the non-resonance case $\omega_s\neq \omega_c$ in Eq.~(\ref{spectrum}). In the non-resonance case, the energy renormalization occurs. In the left panel of Fig.~\ref{non-resonance}, we show the renormalized frequency $\omega^r_c(t)$, the dissipation coefficient $\gamma(t)$, the fluctuation coefficient $\tilde{\gamma}(t)$, and in the right panel, it shows the quantum work power $\mathcal{P}_{w}^{e}(t)$, the dissipation heat current $\mathcal{I}_{h}^{\mathcal{D}}(t)$, and the fluctuation heat current $\mathcal{I}_{h}^{\mathcal{F}}(t)$, respectively. We can see that in addition to heat, intrinsic quantum work also contributes to the change of internal energy, i.e., $dE^r(t)/dt=\mathcal{P}^e_w(t)+\mathcal{I}_h^{\mathcal{D}}(t)+\mathcal{I}_h^{\mathcal{F}}(t)$. 
As shown by Fig.~\ref{non-resonance}(b,c), non-Markovian oscillations of heat currents is still significant for non-resonance case, but the effect is weaker than the resonance case in comparing with the results in Fig.~\ref{resonance}(a,b). The magnitude of dissipation and fluctuations show that the exchange of energy and information is not as dramatic as in the resonance case. Despite this, the behavior of heat currents are similar for both the resonance and non-resonance cases. Furthermore, the magnitudes of the quantum work power and the dissipation heat current have a two-order difference, as shown in Fig.~\ref{non-resonance}(d–f). This indicates that the energy change is dominated by the quantum heat. On the other hand, the work power caused by the energy renormalization of the system is very weak. In other words, the dissipation effect $\gamma(t)$ or the fluctuation effect $\tilde{\gamma}(t)$ is much stronger than the energy renormalization in this hybrid system. 

In comparing to Fig.~\ref{non-resonance}, we show the corresponding results with the weak coupling in Fig.~\ref{non-resonance_weak}. As a result, the renormalized frequency, the dissipation, and fluctuation coefficients in Fig.~\ref{non-resonance_weak}(a-c) all reach steady values without oscillation. Also, in comparison to Fig.~\ref{non-resonance}(a) and (d), the energy renormalization and the quantum work power both are small in two orders, see Fig.~\ref{non-resonance_weak}(a) and (d). In other words, the strong coupling enhances the energy renormalization and the corresponding quantum work power.
On the other hand, same as the results in Fig.~\ref{resonance_weak}(c,d), the dissipation current is negative and the fluctuation heat current is positive. This is the single-direction heat transfers as Markovian processes.

In summary, we demonstrate the dissipation heat current, the fluctuation heat current, and the quantum work power in the hybrid cavity system coupled to spin ensemble in both the strong and weak couplings. We show how the dissipation and the fluctuation dynamics cause the heat currents. Comparing the results in the strong coupling with that in the weak coupling, we find that the heat currents in the strong coupling exhibit strong non-Markovian dynamics. Thus, in a strong coupling, both energy and information flow in and out of the system in an oscillatory fashion. Furthermore, for the non-resonant situation, the cavity system also do the quantum work due to the renormalization of cavity frequency.

\begin{figure}[h]
  \includegraphics[width=8.3cm, height=9cm]{fig4.pdf}
  \caption{The cavity is strongly coupled to the spin ensemble with main frequency $\omega_s=0.998\omega_c$. (a) The renormalized frequency $\Delta \omega^r_c(t)=\omega^r_c(t)-\omega_c$, (b) the dissipation coefficient $\gamma(t)$, (c) the fluctuation coefficient $\tilde{\gamma}(t)$, (d) the quantum work power $\mathcal{P}^{e}_{\omega}(t)$, (e) the dissipation heat current $\mathcal{I}_{h}^{\mathcal{D}}(t)$, and (f) the fluctuation heat current $\mathcal{I}_{h}^{\mathcal{F}}(t)$ are plotted as a function of time. In (d-f), the units are $\hbar \omega_c/ns$. The system is initially prepared in a coherent state $|z_0\ket$ with $z_0=10$.  Other parameters are the coupling strength $\Omega=17.2\pi$ MHz, the decay constant $\kappa=0.8\pi$ MHz, and the temperature $T=0.1$~K.}  
  \label{non-resonance}
\end{figure}

\begin{figure}[h]
  \includegraphics[width=8.4cm, height=9cm]{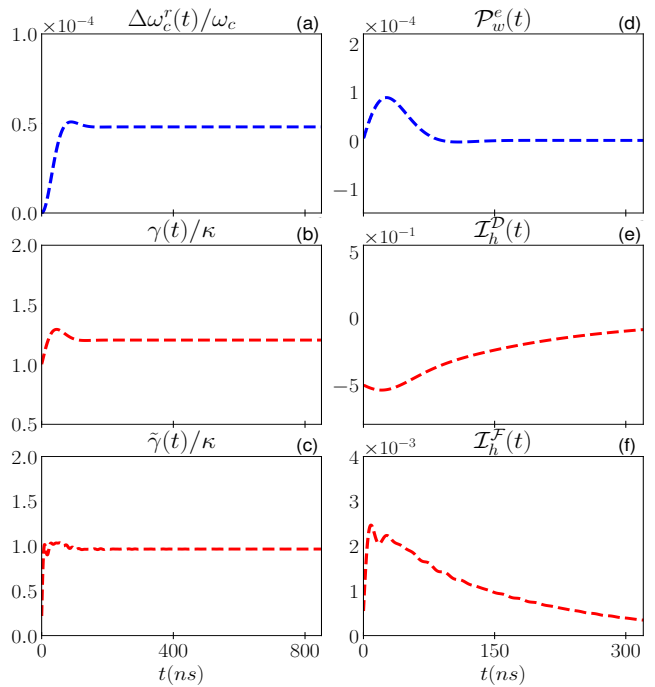}
  \caption{The cavity is weakly coupled to the spin ensemble with main frequency $\omega_s=0.998\omega_c$. (a) The renormalized frequency $\Delta \omega^r_c(t)=\omega^r_c(t)-\omega_c$, (b) the dissipation coefficient $\gamma(t)$, (c) the fluctuation coefficient $\tilde{\gamma}(t)$, (d) the quantum work power $\mathcal{P}^{e}_{\omega}(t)$, (e) the dissipation heat current $\mathcal{I}_{h}^{\mathcal{D}}(t)$, and (f) the fluctuation heat current $\mathcal{I}_{h}^{\mathcal{F}}(t)$ are plotted as a function of time. The coupling strength $\Omega$ is $1.72\pi$ MHz. Other parameters and units are the same in Fig.~\ref{non-resonance}.}  
  \label{non-resonance_weak}
\end{figure}

\subsubsection{The transient energy exchanges due to non-Markovian dynamics}
Next, we consider the transient thermodynamics of the system under driving. The system is prepared in a vacuum state and applied by an oscillating driving field $f(t)=f_me^{-i\omega_c}\theta(t-t_s)$, where $\omega_c$ is the cavity photon frequency and $\theta(t-t_s)$ means that the driving is applied only in the time interval $t_0\rightarrow t_s$. Because the renormalization of energy is very weak as shown in Fig.~\ref{non-resonance}, the total quantum work power is dominated by the driving. On the other hand, the fluctuation heat current does not depend on the driving field. At low temperature, the fluctuation heat current  $\mathcal{I}_h^\mathcal{F}(t)$ is much smaller than the dissipation heat current  $\mathcal{I}_h^\mathcal{D}(t)$, as shown in Fig.~\ref{resonance}(d). That is, the total energy change is mainly determined by $dE^r(t)/dt\approx\mathcal{P}_w^d(t)+\mathcal{I}_h^D(t)$. In Fig.~\ref{non_res}(a-f), we show the driving field amplitude $\tilde{f}(t)$, the driving induced cavity field amplitude $\tilde{y}(t)$, the renormalized driving field amplitude $\tilde{f_r}(t)$ due to the cavity coupling with the spin ensemble, and also the internal energy of the cavity $E^r(t)$, the work power by the driving $\mathcal{P}_w(t)$, as well as the heat current $\mathcal{I}_h(t)$, respectively. 

Applying the driving $f(t)$ as shown Fig.~\ref{non_res}(a), the cavity field, the renormalized driving field, and the internal energy exhibit the oscillating behavior at the beginning, see Fig.~\ref{non_res}(b-d). From Eqs.~(\ref{coeff_MEQ}b) and (\ref{equyv}b), one can find that the cavity field $y(t)$ and the renormalization of the driving field $f_r(t)$ are both affected by the non-Markovian cavity dynamics due to the cavity coupling to the spin ensemble. The internal energy thus also exhibits the oscillating behavior. Consequently, the heat current and the work power are significantly affected by the non-Markovian effects, as shown in Fig.~\ref{non_res}(e,f). Notably, the work power done by the renormalized driving field $f_r(t)$ could be negative due to the non-Markovian effect, see Eqs.~\ref{coeff_MEQ}(b) and \ref{WQ_MEQ}(a). On the other hand, after a certain time, both the cavity field and the renormalized field both reach to steady states. This is because the external driving energy reaches a balance with the dissipation energy of the system. At $t=t_s$, we turn off the driving. The cavity field suddenly decays but meantime the energy retrieves back from the spin ensemble. This results in the oscillation of the cavity field and the internal energy again after $t_s$, as shown in Fig.~\ref{non_res}(b,d).

\begin{figure}[h]
 \includegraphics[width=8.4cm, height=9cm]{Fig6.pdf}
 \caption{ A driving $f(t)=f_me^{-i\omega_ct}\theta(t-t_s)$ is applied to the cavity which is strongly coupled to the spin ensemble with main frequency $\omega_s=0.998\omega_c$. (a) The driving amplitude $\tilde{f}(t)=f(t)e^{i\omega_ct}$, (b) the induced cavity field amplitude $\tilde{y}(t)=y(t)e^{i\omega_ct}$, (c) the renormalized field amplitude $\tilde{f_r}(t)=f_r(t)e^{i\omega_ct}$, where $y(t)$ and $f_r(t)$ are determined by Eqs.~(\ref{coeff_MEQ}b) and (\ref{equyv}b), (d) the internal energy $E^r(t)$, (e) the quantum work power $\mathcal{P}_w(t)$ (blue-solid) and (f) the quantum heat current $\mathcal{I}_h(t)$ (red-dash) are plotted as a function of time. In (a-c), the blue-solid and orange-dash lines correspond to the real and the image part, respectively. The units in (d-f) are $\hbar \omega_c/ns$. The system is initially prepared in the vacuum state $|0\ket$. Other parameters are the driving amplitude $f_m=\hbar \omega_c/10$, the turn off time $t_s=900$(ns), the coupling strength $\Omega=17.2\pi$ MHz, the decay constant $\kappa=0.8\pi$ MHz, and the temperature $T_0=0.1$~K.}\label{non_res}
\end{figure}

\begin{figure}[h]
  \includegraphics[width=8.4cm, height=6.8cm]{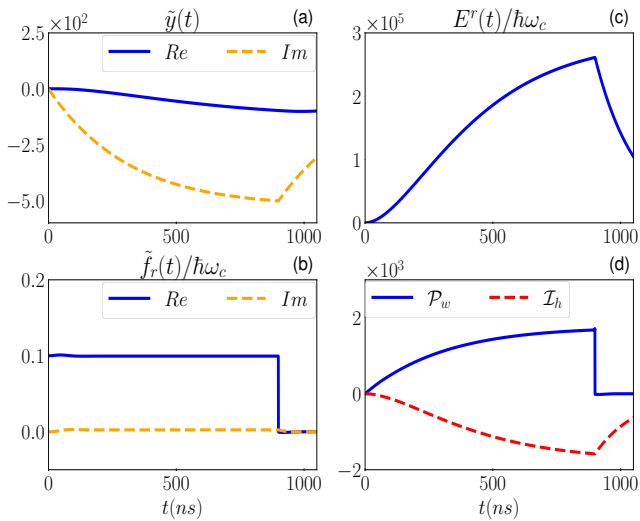}
  \caption{A driving $f(t)=f_me^{-i\omega_ct}\theta(t-t_s)$ is applied to the cavity which is weakly coupled to the spin ensemble with main frequency $\omega_s=0.998\omega_c$. (a) The induced cavity field amplitude $\tilde{y}(t)=y(t)e^{i\omega_ct}$, (b) the renormalized field amplitude $\tilde{f_r}(t)=f_r(t)e^{i\omega_ct}$, (c) the internal energy $E^r(t)$, (d) the quantum work power $\mathcal{P}_w(t)$ (blue-solid) and the quantum heat current $\mathcal{I}_h(t)$ (red-dash) are plotted as a function of time. The coupling strength  $\Omega=1.72\pi$ MHz. Other parameters and units are the same in Fig.~\ref{non_res}.}
  \label{non_res_weak}
\end{figure}

As a comparison, we show the corresponding results in the weak coupling regime in Fig.~\ref{non_res_weak}. Because of the Markovian dynamics, in contrast to Fig.~\ref{non_res}, there is no oscillation for all quantities shown in Fig.~\ref{non_res_weak}. The flowing-back energy from the spin ensemble is also negligible when the driving field turned off. Furthermore, the renormalized driving field is the same as the original driving field, as shown in Fig.~\ref{non_res_weak}(b). In other words, the renormalization effect is negligible in the driving field due to the weak coupling. 
On the other hand, Comparing Fig.~\ref{non_res}(e) with Fig.~\ref{non_res_weak}(d), we notice that when the system is more weakly coupled to the spin ensemble, the driving-induced quantum work power is stronger because the dissipation is weaker in the weak coupling regime. This indicates that the enhancement of dissipation reduces the driving-induced work power. As a result, the quantum work power in the strong coupling system is weaker, even for the same driving source.

To get a more complete picture for the difference between the strong and weak couplings, we consider the system with a resonant driving field, i.e., we turn the frequency of the driving field to the steady value of the renormalized frequency $\omega^r_c$. In Fig.~\ref{drive-resonace}, we show the driving field amplitude $\tilde{f}(t)$, the driving induced cavity field amplitude $\tilde{y}(t)$, the renormalized driving field amplitude $\tilde{f_r}(t)$, the internal energy $E^r(t)$, the work power $\mathcal{P}_w(t)$, and the heat current $\mathcal{I}_h(t)$, respectively, in the strong coupling. In Fig.~\ref{drive-resonace}(a), the induced cavity field monotonically increases and its amplitude is larger in comparing with the non-resonance driving case, see Fig.~\ref{non_res}(a). Similar to Fig.~\ref{non_res}(b), Fig.~\ref{drive-resonace}(b) shows that the renormalized driving field exhibits non-Markovian oscillates at the beginning and eventually reaches the steady state. Furthermore, although the strong coupling reduces the renormalized driving field, the steady value of the renormalized driving field in the resonance case is only slightly smaller than the original driving field, in comparing with the non-resonance one, see Fig.~\ref{non_res}(c). Meanwhile, the heat current and work power also exhibit stronger non-Markovian oscillations as shown in Fig.~\ref{drive-resonace}(e,f). That is, the larger magnitudes of the work power and the heat current demonstrate that energy and information flow into and out of the system quickly, compared to the non-resonant results in Fig.~\ref{non_res}(d). In Fig.~\ref{drive-resonace}(e) it also shows the field oscillation after the driving is turned off. It is worth noting that by comparing to Fig.~\ref{non_res}(e,f), we find that the work power and the heat current in the resonant drive system have only one-frequency oscillations, as a consequence of resonance driving. On the other hand, in the weak coupling regime, the renormalized effect is almost negligible so that thermodynamics properties remain unchanged, just as shown in Fig.~\ref{non_res_weak}. Furthermore, the results in Figs.~\ref{non_res} and \ref{drive-resonace} show that although the frequency renormalization is not significantly large, the work power is very sensitive to the driving frequency, as a flexible controllability in practical application.

From the above discussion, we find that thermodynamics quantities, especially the quantum work power and the quantum heat currents, are so sensitive with the non-Markovian effect. In the last, we shall study the strong coupling system whose frequency equals to the main frequency of the spin ensemble, i.e., $\omega_c=\omega_s$. In this situation, no frequency shift (no frequency renormalization) occurs even in the strong coupling. In Fig.~\ref{tri-resonance}, we show the driving field amplitude $\tilde{f}(t)=f(t)e^{i\omega_ct}$, the induced cavity field amplitude $\tilde{y}(t)=y(t)e^{i\omega_ct}$, the renormalized driving field amplitude $\tilde{f}_r(t)=f_r(t)e^{i\omega_ct}$, the internal energy $E^r(t)$, the work power $\mathcal{P}_w$, and the heat current $\mathcal{I}_h$, respectively, for the resonant driving, i.e.,  $\omega_s=\omega_c=\omega_{d}$. Figure~\ref{tri-resonance}(b) shows the behavior of the cavity field, which is similar to the solution given in Fig.~\ref{non_res}(b). But only the imaginary part of filed amplitude exists, i.e., there is a $\pi/2$-phase shift between the induced cavity field $y(t)$ and the original driving field $f(t)$. This is because  $u(t,\tau)f(\tau)$ in Eq.~(\ref{y}) is real. The $\pi/2$-phase shift comes purely from the front factor $\f{1}{i\hbar}$ in Eq.~(\ref{y}). With the same reason, the renormalized driving field only has a real amplitude and thus has the same phase with the original driving field, as shown in Fig.~\ref{tri-resonance}(c). On the other hand, we can see that in Fig.~\ref{tri-resonance}(d), the magnitude of the internal energy is smaller in comparing with the result in Fig.~\ref{drive-resonace}(d) for $\omega_d=\omega_s\neq\omega_c$. This is because the resonance between the system and the spin ensemble makes stronger dissipation, as shown in Fig.~\ref{resonance}(a). In more detail, Fig.~\ref{tri-resonance}(e,d) show that the quantum work power and the quantum heat current exhibit fast and dramatic energy exchange due to the strong non-Markovian effect in comparison. Finally, we also study the system in the weak coupling regime with the same other conditions and find that the results are similar with the results as shown in Fig.\ref{non_res_weak} because of the negligible non-Markovian effects.

\begin{figure}[h]
  \includegraphics[width=8.4cm, height=9cm]{Fig8.pdf}
  \caption{A driving $f(t)=f_me^{-i\omega^r_ct}\theta(t-t_s)$ is applied to the cavity which is strongly coupled to the spin ensemble with main frequency $\omega_s=0.998\omega_c$. (a) The driving $\tilde{f}(t)=f(t)e^{i\omega^r_ct}$, (b) the induced cavity  field $\tilde{y}(t)=y(t)e^{i\omega^r_ct}$, (c) the renormalized field $\tilde{f_r}(t)=f_r(t)e^{i\omega^r_ct}$, where $y(t)$ and $f_r(t)$ are determined by Eqs.~(\ref{coeff_MEQ}b) and (\ref{equyv}b), (d) the internal energy $E^r(t)$, (e) the quantum work power $\mathcal{P}_w(t)$ and (f) the quantum heat current $\mathcal{I}_h(t)$ are plotted as a function of time. Other parameters and units are the same in Fig.~\ref{non_res}.}
  \label{drive-resonace}
\end{figure}

\begin{figure}[h]
  \includegraphics[width=8.4cm, height=9cm]{Fig9.pdf}
  \caption{A driving $f(t)=f_me^{-i\omega_ct}\theta(t-t_s)$ is applied to the cavity which is strongly coupled to the spin ensemble with main frequency $\omega_s=\omega_c$. (a) The driving $\tilde{f}(t)=f(t)e^{i\omega_ct}$, (b) the induced cavity  field $\tilde{y}(t)=y(t)e^{i\omega_ct}$, (c) the renormalized field $\tilde{f_r}(t)=f_r(t)e^{i\omega_ct}$, where $y(t)$ and $f_r(t)$ are determined by Eqs.~(\ref{coeff_MEQ}b) and (\ref{equyv}b), (d) the internal energy $E^r(t)$, (e) the quantum work power $\mathcal{P}_w(t)$ and (f) the quantum heat current $\mathcal{I}_h(t)$ are plotted as a function of time. Other parameters and units are the the same in Fig.~\ref{non_res}.}\label{tri-resonance}
\end{figure}

\section{conclusions and perspectives}\label{conclusion}

We apply the renormalization theory of quantum thermodynamics we developed recently to investigate transient quantum work power and quantum heat current in a strong-coupling system with its environment.  The quantum work arises from the energy renormalization and the renormalized driving field. Through the renormalization of energy levels, the system and the environment do the quantum work on each other. On the other hand, the driving-induced system work power involves the non-Markovian effect due to the feedback of the non-Markovian dynamics of the system. Furthermore, the total heat current consists of the dissipation heat current and the fluctuation heat current, which purely arise from the dissipation and the thermal fluctuations dynamics of the open system due to strong coupling with its environnement.

We study further the strong-coupling hybrid system under driving , which consists of a superconducting microwave cavity coupled with a spin ensemble of NV centers in diamond. With the controllability of the coupling in experiments, one can tune the coupling strength and the main frequency of the spin ensemble to examine the transient energy conversion. We find that the strong coupling between the cavity and the spin ensemble induces the strong non-Markovian memory effects on energy renormalization, and the corresponding quantum work power. On the other hand, the frequency shift (renormalization) is relatively small in this strong coupling system, but the driving field is very sensitive when it is resonant with the cavity. The driving-induced work power is thus enhanced significantly in the resonance system. Furthermore, we can tune the system coupling to manipulate the driving-induced quantum work power through the renormalization and non-Markovian effects. There provides a new avenue in studying the non-Markovian transient quantum heat and work through quantum engineering, and developing strong-coupling quantum engines.

\end{document}